\documentclass[journal]{IEEEtran}
\usepackage{graphicx}
\usepackage{amsmath,amsfonts,amssymb,amsthm}
\usepackage{bm}
\usepackage{booktabs}
\usepackage{tabularx}
\usepackage{multirow}
\usepackage{cite}
\usepackage{algorithm}
\usepackage{algpseudocode}
\usepackage{url}
\usepackage[T1]{fontenc}
\usepackage{microtype}

\newtheorem{proposition}{Proposition}
\theoremstyle{remark}
\newtheorem{remark}{Remark}

\newcommand{\Ybus}{\bm{Y}_{\mathrm{bus}}}
\newcommand{\Vbus}{\bm{V}}
\newcommand{\Sbus}{\bm{S}_{\mathrm{bus}}}

\newcommand{\nnz}{\operatorname{nnz}}

\begin{document}

\title{Symbolic Direct Construction of Augmented Systems for Levenberg--Marquardt Power Flow}

\author{Tianshi~Cheng,~\IEEEmembership{Member,~IEEE,}
        and~Venkata~Dinavahi,~\IEEEmembership{Fellow,~IEEE}
\thanks{T.~Cheng is with RTDS Technologies Inc., Winnipeg, MB R3T~2E1, Canada, and V.~Dinavahi is with the Department of Electrical and Computer Engineering, University of Alberta, Edmonton, AB T6G~2V4, Canada (e-mail: tcheng1@ualberta.ca; dinavahi@ualberta.ca).}}

\maketitle

\begin{abstract}
Levenberg--Marquardt power flow (LMPF) addresses ill-conditioned equations and provides least-squares stationary points for infeasible operating requests. Conventional implementations form damped normal equations, where $J^{\top}J$ introduces fill, squares the spectral condition number of the undamped part, and adds sparse-product cost. We propose an augmented LMPF formulation with a symbolic-permutation-based Jacobian operator that fills the augmented blocks directly. We prove that the matrix is symmetric quasi-definite for positive damping, enabling fixed-order, pivot-free QDLDL factorization. This property also motivates the solver choice: KLU value-only refactorization can reuse an unsuitable pivot order, forcing fresh full factorizations on difficult LMPF trajectories. We further introduce one-probe adaptive trust-region initialization for the radius and damping.

Direct assembly is $7.5$--$11.9\times$ faster than the fastest coordinate-list path and $18.0$--$58.8\times$ faster than cached normal-equation sparse-product formation; QDLDL reduces numerical refactorization and triangular-solve time by up to $3.40\times$ on identical matrices. Adaptive initialization solves flat-started RTE cases in 19--25 factorizations, $7.5$--$11.6\times$ faster than reproduced three-step LM. On case2383wp, our 84-iteration run performs 84 factorizations and 168 resolves in 237~ms, versus published times of 200.780--441.72~s with fewer outer iterations, corresponding to speedups of $847$--$1860\times$. These results require complete factorization and resolve accounting and reveal initialization and legacy-parameter effects. DC-initialized NR exhibits fast quadratic convergence throughout every tested solvable loading range, motivating clearer, reproducible definitions of ill-conditioned power-flow benchmarks.
\end{abstract}


\section{Introduction}
\label{sec:introduction}
\IEEEPARstart{N}{ewton--Raphson} (NR) power flow is highly efficient within its convergence basin, but its behavior under poor initialization or near a singular Jacobian remains a persistent research topic. Iwamoto and Tamura introduced an optimal-multiplier treatment for ill-conditioned systems~\cite{iwamoto1981}, while Tripathy \emph{et al.} developed a Newton-like alternative~\cite{tripathy1982}. Levenberg--Marquardt (LM) regularization~\cite{levenberg1944,marquardt1963} has since been applied to power flow through damping, trust-region, and composite-step variants~\cite{lagace2008,derakhshandeh2016,derakhshandeh2018,pires2019}.

Three separate issues are often conflated in these investigations. The first is \emph{numerical initialization}: a solution may exist, yet the flat-start voltage may lie outside the NR convergence basin. The second is \emph{existence}: beyond a saddle-node bifurcation, the specified loading direction has no real power-flow solution. The third is \emph{near-singularity}: the Jacobian becomes ill-conditioned near a fold, even when a solution remains reachable. Reliable attribution of convergence or failure requires comparisons that separate these conditions.

The implementation introduces a further confound. LM iterations form a damped least-squares system, and a single accepted outer step may require several trial factorizations. Outer-iteration counts therefore miss a substantial and highly variable fraction of the work. Moreover, explicitly forming $J^{\top}J$ introduces a sparse matrix product, while conventional coordinate (COO) assembly and transpose operations add allocation, indirect addressing, and memory traffic. These overheads can obscure both the cost of the numerical method and the effect of the sparse factorization backend.

This paper extends our symbolic direct Jacobian construction~\cite{cheng2026symbolic}, implemented in the open-source \texttt{RustPower} solver~\cite{rustpower}, to the augmented LM system. It combines direct sparse construction, a verified symmetric solver class, adaptive step initialization, and complete cost accounting.

The contributions are as follows.
\begin{enumerate}
    \item \textbf{An augmented LM formulation with a verified solver class.} We establish exact equivalence with the damped normal equations, derive the singular-value filtering limits, and prove quasi-definiteness and inertia for every positive damping value and every Jacobian. The analysis also gives the positive damping floor required by pivot-free LDL$^{\top}$ factorization.
    \item \textbf{Symbolic direct construction from $\Ybus$.} Sorted admittance columns yield the reduced Jacobian pattern by prefix selection. Base addresses and segment lengths determine every output-column base, diagonal rank, and quadrant offset inside the fill loop. This integer address arithmetic replaces auxiliary offset tables, COO triplets, runtime sorting, and transpose scatter. A row-oriented upper-triangle path nearly halves stored augmented-matrix nonzeros.
    \item \textbf{Controlled implementation comparisons.} Assembly variants execute identical iterates, backend comparisons use identical augmented matrices, and robustness comparisons count every factorization and triangular resolve. Direct assembly achieves $7.5$--$11.9\times$ speedups over the fastest COO path and $18.0$--$58.8\times$ over cached normal-equation sparse-product formation. The respective end-to-end speedups are $1.15$--$1.43\times$ and $1.28$--$3.57\times$. QDLDL reduces numerical refactorization plus triangular-solve time by up to $3.40\times$ relative to KLU.
    \item \textbf{Adaptive step initialization and reproducible evaluation.} Our one-probe extension jointly initializes the trust-region radius and damping scale. It solves the flat-started RTE cases in 19--25 factorizations and is $7.5$--$11.6\times$ faster than the reproduced three-step LM. Full accounting exposes nested work that outer-iteration counts miss. The case2383wp and RTE experiments separate initialization, existence, and near-singularity, providing concrete requirements for reproducible ill-conditioned power-flow benchmarks.
\end{enumerate}

Section~\ref{sec:formulation} defines the power-flow and LM equations. Section~\ref{sec:augmented} develops the augmented quasi-definite system. Section~\ref{sec:assembly} derives the direct-fill construction. Section~\ref{sec:protocol} specifies step control and experimental accounting, Section~\ref{sec:results} reports results, and Section~\ref{sec:discussion} discusses the appropriate role of LM power flow.

\section{Power-Flow and LM Formulations}
\label{sec:formulation}

\subsection{Reduced Power-Flow Equations}
Buses are ordered as $[\mathrm{PQ}\mid\mathrm{PV}\mid\mathrm{slack}]$. Let $n_{pq}$ and $n_{pv}$ denote their counts, and define $n_a=n_{pq}+n_{pv}$ and $n=n_{pv}+2n_{pq}$. The retained polar state is
\begin{equation}
    x=[\theta_0,\ldots,\theta_{n_a-1},
       |V_0|,\ldots,|V_{n_{pq}-1}|]^{\top}.
\end{equation}
The mismatch vector $F(x)\in\mathbb{R}^{n}$ contains active-power mismatches at all nonslack buses and reactive-power mismatches at PQ buses. With $J=\partial F/\partial x$, the NR correction solves
\begin{equation}
    J\delta=-F.
    \label{eq:newton}
\end{equation}
The numerical Jacobian is evaluated from $\Sbus=\Vbus\odot\operatorname{conj}(\Ybus\Vbus)$ using the same complex-domain expressions as the production NR solver.

\subsection{Least-Squares and Damped Steps}
LM methods replace the root equation with minimization of
\begin{equation}
    f(x)=\tfrac{1}{2}\|F(x)\|_2^2.
    \label{eq:merit}
\end{equation}
A damped Gauss--Newton step satisfies
\begin{equation}
    (J^{\top}J+\mu D)\delta=-J^{\top}F,
    \label{eq:damped_normal}
\end{equation}
where $\mu>0$ and $D\succ0$ is diagonal. The experiments use $D=I$ in the retained polar coordinates; the implementation also supports voltage- and network-dependent diagonal metrics.

\subsection{Singular-Value Filtering and Step Limits}
For $D=I$, let $J=U\Sigma V^{\top}$ be a full singular-value decomposition. The damped step is
\begin{equation}
    \delta(\mu)=-\sum_{i=1}^{n}
    \frac{\sigma_i}{\sigma_i^2+\mu}(u_i^{\top}F)v_i,
    \label{eq:svd_filter}
\end{equation}
with the convention $\sigma_i/(\sigma_i^2+\mu)=0$ when $\sigma_i=0$ and $\mu>0$. Thus damping filters each singular direction differently: well-observed directions retain their Newton correction, whereas small singular values are strongly attenuated. In particular,
\begin{equation}
    \lim_{\mu\to0^+}\delta(\mu)=-J^{\dagger}F.
    \label{eq:mn_limit}
\end{equation}
For full-column-rank $J$, this is the Newton correction~\eqref{eq:newton}; at a rank-deficient point, it is the minimum-norm least-squares correction. For general positive diagonal $D$, the same statement holds after scaling the step by $D^{1/2}$. At the opposite extreme,
\begin{equation}
    \delta(\mu)=-\frac{1}{\mu}D^{-1}J^{\top}F+O(\mu^{-2}),
    \label{eq:gradient_limit}
\end{equation}
so large damping approaches a scaled steepest-descent direction. These limits describe the damping-controlled direction at a fixed initial point.

A trial is assessed using the actual-to-predicted reduction ratio
\begin{equation}
    \rho=\frac{f(x)-f(x+\delta)}
    {\tfrac{1}{2}\left(\mu\delta^{\top}D\delta-(J^{\top}F)^{\top}\delta\right)}.
    \label{eq:gain_ratio}
\end{equation}
The denominator is the decrease of the undamped quadratic model evaluated at the damped step. Using~\eqref{eq:damped_normal}, it also equals $\tfrac12\|J\delta\|_2^2+\mu\delta^{\top}D\delta$. The damping update follows the classical Nielsen rules~\cite{nielsen1999}: accept a step with $\rho>10^{-4}$, reduce $\mu$ after a high-quality step, and increase it after rejection. Section~\ref{sec:protocol} defines two trust-region methods: TR with a prescribed initial radius, and adaptive TR with a probe-initialized radius and damping. An accepted outer iteration can contain several trial factorizations, whereas composite steps can reuse one factorization with several right-hand sides.

\subsection{DC Initialization}
The dc power-flow approximation supplies an angle initial estimate through one linear solve. Voltage magnitudes remain at their specified flat-start values. Phase-shifting transformers enter through their associated injection corrections; neglecting these terms can change the quality of the initial point substantially. We count the dc linear solve as initialization and report subsequent NR iterations separately.

\section{The Augmented Quasi-Definite System}
\label{sec:augmented}

\subsection{Why Avoid Explicit Normal-Equation Formation}
For $\mu>0$, $J^{\top}J+\mu D$ is positive definite and appears suitable for sparse Cholesky factorization. Three practical difficulties motivate a different formulation.

First, explicit formation of $J^{\top}J$ requires a sparse matrix product and can introduce fill before the linear solver is invoked. It also squares the spectral condition number of $J$ in the undamped part. This is particularly undesirable near a fold, where recovering a Newton-like direction requires small $\mu$.

Second, general sparse LU with numerical pivoting provides a different reuse contract from a fixed-pattern symmetric factorization. KLU value-only refactorization reuses an earlier pivot order, which can become unsuitable as the LMPF iterates change. Our flat-started case6515rte control stalls without reporting a factorization failure, whereas fresh factorization of the same matrices converges. A reliable KLU path must then perform a fresh full factorization at every trial on such trajectories. This reliability requirement and its full-factorization cost are primary motivations for QDLDL, alongside the measured speed difference.

Third, the normal-equation product discards the assembly advantage already available in the direct Jacobian kernel. The augmented formulation avoids this product entirely.

\subsection{Exact Augmented Form}
Consider
\begin{equation}
    \underbrace{\begin{bmatrix}
        \mu D & J^{\top}\\
        J & -I
    \end{bmatrix}}_{K_\mu}
    \begin{bmatrix}\delta\\ \lambda\end{bmatrix}
    =
    \begin{bmatrix}0\\-F\end{bmatrix}.
    \label{eq:augmented}
\end{equation}
The second block equation gives $\lambda=J\delta+F$. Substitution into the first yields
\begin{equation}
    \mu D\delta+J^{\top}(J\delta+F)=0,
\end{equation}
which is exactly~\eqref{eq:damped_normal}. Thus~\eqref{eq:augmented} preserves the equations and the iterate while changing the sparse linear-system representation.

\begin{proposition}[Quasi-definiteness and inertia for positive damping]
\label{prop:quasi_definite}
If $\mu>0$ and $D$ is positive diagonal, then $K_\mu$ in~\eqref{eq:augmented} is symmetric quasi-definite for every real square matrix $J$, and $\operatorname{inertia}(K_\mu)=(n,n,0)$.
\end{proposition}
\begin{proof}
The leading block satisfies $\mu D\succ0$, and the trailing block satisfies $-I\prec0$. Hence $K_\mu$ has the block form
\[
\begin{bmatrix}A&B^{\top}\\B&-C\end{bmatrix},
\qquad A\succ0,\quad C\succ0,
\]
which is quasi-definite in the sense of Vanderbei~\cite{vanderbei1995}. Moreover, block elimination by congruence gives
\begin{align}
E_\mu K_\mu E_\mu^{\top}&=
\begin{bmatrix}
\mu D&0\\
0&-I-J(\mu D)^{-1}J^{\top}
\end{bmatrix},
\label{eq:inertia_congruence}\\
E_\mu&=\begin{bmatrix}
I&0\\-J(\mu D)^{-1}&I
\end{bmatrix}.\nonumber
\end{align}
The Schur complement is negative definite for every $J$. Sylvester's law therefore gives $n$ positive and $n$ negative eigenvalues and no zero eigenvalue.
\end{proof}

Every symmetric permutation of a quasi-definite matrix admits an LDL$^{\top}$ factorization with nonzero diagonal pivots. A fill-reducing symbolic ordering can consequently be computed once and reused for all Jacobian values and all positive $\mu$, without relying on numerical pivot changes to preserve a nonzero-pivot sequence. QDLDL consumes the upper triangle and performs this quasi-definite LDL$^{\top}$ factorization~\cite{osqp2020}. KLU requires the full augmented matrix and does not exploit its symmetry~\cite{davis2010klu}.

\begin{remark}[Positive damping floor]
At $\mu=0$, the leading block loses positive definiteness, so the quasi-definite proof requires $\mu>0$. A small positive floor $\mu_{\min}>0$ therefore belongs to both the solver contract and step control. When the floor is sufficiently small, the computed step is numerically close to the undamped Gauss--Newton direction while the augmented matrix remains in the required class.
\end{remark}

\begin{remark}[Role of an identity shift]
Replacing $\mu D$ by $I+\mu D$ would preserve quasi-definiteness at $\mu=0$, but block elimination would produce $J^{\top}J+I+\mu D$. The permanent identity shift would survive as $\mu\to0$ and alter the undamped Gauss--Newton equation. The implemented damping block therefore vanishes with $\mu$, subject to the positive numerical floor.
\end{remark}

\subsection{Damping-Only Updates}
In~\eqref{eq:augmented}, $\mu$ appears only in the leading diagonal. After one assembly, a damping retry updates $n$ scalar slots and invokes numerical refactorization while reusing $J$, $J^{\top}$, the $-I$ block, and the sparsity pattern. Composite methods can likewise invoke triangular solves with additional right-hand sides while retaining the same factorization. These operations make the number of factorizations and resolves directly observable in the implementation.

\section{Symbolic Direct Construction}
\label{sec:assembly}

\subsection{Admittance-Prefix Structure}
Let the pre-permuted admittance matrix be stored in CSC form by column offsets $p^{Y}_j$, row indices $r^{Y}$, and values $y$. Rows are sorted within each column. For column $j$, define
\begin{align}
    \pi_j&=\#\{i:Y_{ij}\ne0,\ i<n_{pq}\},\nonumber\\
    \alpha_j&=\#\{i:Y_{ij}\ne0,\ i<n_a\},
    \label{eq:segment_cuts}
\end{align}
and let $d_j$ be the local rank of the diagonal entry within that column. These three quantities are obtained during symbolic analysis.

An angle column of the reduced Jacobian contains active-power rows followed by reactive-power rows. Its two row segments are therefore prefixes of lengths $\alpha_j$ and $\pi_j$ of the corresponding $\Ybus$ column. A magnitude column exists only for a PQ bus and uses the same prefixes with the block roles exchanged. Hence the reduced Jacobian pattern is a type-restricted view of the existing admittance graph.

Let $c^{J}_j$ denote the CSC start of reduced-Jacobian column $j$, and let $c^{\mathsf T}_r$ denote the corresponding start in its structural transpose. Both are symbolic prefix sums of retained segment lengths; the latter is equivalently a prefix sum of reduced-Jacobian row lengths. The complete augmented pattern follows by placing these graphs in the off-diagonal blocks and appending the two diagonal blocks. No numerical values are required for this construction.

\subsection{Addresses Derived Inside the Fill Loop}
The implementation separates a block base from the local offset within that block. At the beginning of each output-column or output-row iteration, it reads the relevant $\Ybus$ start, the retained segment lengths $\pi_j$ and $\alpha_j$, and the diagonal rank $d_j$. It then derives the destination addresses immediately before filling that column or row.

For example, in the upper-triangle layout of~\eqref{eq:augmented}, a state column $r<n$ contains only its damping diagonal and therefore starts at address $r$. Residual column $n+r$ starts at
\begin{equation}
    b^{K}_{n+r}=n+c^{\mathsf T}_r+r,
    \label{eq:upper_base}
\end{equation}
because each preceding residual column contributes a row segment of the transposed Jacobian and one $-I$ entry. The $-I$ entry of the current column is the last entry before the next column start. If $d_r$ is the local diagonal rank in the retained row segment, the diagonal correction is applied at $b^{K}_{n+r}+d_r$ rather than located by a search. Thus the global column base, the start of each retained segment, and the diagonal address are all obtained by integer addition at loop entry.

The column-oriented full-layout path applies the same principle to all four Jacobian quadrants. If $b$ is the base of the current output block, then the two retained segments begin at $b$ and $b+\alpha_j$; the paired quadrant begins at its own derived base and uses the same local offsets. Each loop forms the starts required by its current column, replacing persistent $J_{11}$, $J_{21}$, $J_{12}$, and $J_{22}$ start arrays.

Shared graph bases and segment lengths keep the global layout and local writes consistent by construction. A few integer additions on values already used by the fill loop replace dependent loads and offset-table maintenance.

\subsection{Column Fill and Direct Transpose Fill}
A column-oriented kernel evaluates each retained admittance entry and writes the corresponding Jacobian values directly into their final positions. Bus-level quantities are hoisted out of the inner loop, and diagonal corrections are applied once at rank $d_j$. For the full augmented form, $J$ and $J^{\top}$ may be emitted together; recomputing a small number of shared products is preferable to a scattered transpose-copy pass when memory traffic dominates.

For the upper-triangle QDLDL path, only $J^{\top}$ is required. The row-oriented kernel computes Jacobian row $r$ directly into residual column $n+r$ of $K_\mu$. A symbolic transpose or mirror map supplies the admittance entries in row order, so values are written contiguously without materializing a separate numerical $J$ block.

Phase-shifting transformers make $\Ybus$ numerically unsymmetric, but each branch still contributes the same off-diagonal pattern in both directions. Thus the \emph{pattern} remains symmetric and the prefix cuts and diagonal ranks apply to the row walk unchanged. Numerical admittance values are always read from the correct directed entry.

Algorithm~\ref{alg:direct_fill} summarizes the resulting organization. Diagonal and off-diagonal blocks are filled in separate single-purpose sweeps, while each column or row derives its own addresses on entry.

\begin{algorithm}[t]
\caption{Direct upper-triangle augmented fill}
\label{alg:direct_fill}
\begin{algorithmic}[1]
\Require Pre-permuted $\Ybus$, symbolic cuts $\pi,\alpha,d$, graph starts $c^{\mathsf T}$, voltage and calculated-power vectors
\For{$j=0,\ldots,n-1$}
    \State $b\gets j$; values$[b]\gets\mu D_{jj}$
\EndFor
\For{each retained residual column $n+r$}
    \State $b\gets n+c^{\mathsf T}_r+r$ \Comment{Eq.~\eqref{eq:upper_base}}
    \State Derive $\pi_r,\alpha_r,d_r$ from the symbolic admittance view
    \For{each retained admittance entry in Jacobian row $r$}
        \State Evaluate the required $J^{\top}$ entry
        \State Write it at $b$ plus its local segment offset
    \EndFor
    \State Apply the diagonal correction at $b+d_r$
    \State Write $-1$ at the final slot of column $n+r$
\EndFor
\end{algorithmic}
\end{algorithm}

\subsection{Storage and Retry Complexity}
The full augmented pattern stores $2\nnz(J)+2n$ nonzeros, whereas the upper-triangle pattern stores $\nnz(J)+2n$. The latter is the direct input required by QDLDL. LU backends receive the full layout; triangle-only storage is selected by the backend contract, not by a runtime format conversion.

For $M=\nnz(J)$ retained entries, numerical fill is $\Theta(M)$ and performs no sorting or dynamic structure insertion. A $\mu$ retry changes $n$ diagonal values. Symbolic analysis is performed once per topology and bus-type partition, and subsequent operating-point changes reuse the complete pattern and the factorization ordering.

\section{Step Control and Evaluation Protocol}
\label{sec:protocol}

\subsection{Trust-Region LM}
Trust-region LM combines the damped step~\eqref{eq:damped_normal} with a bound $\|\delta\|_D\leq\Delta$. We call the method \textbf{TR} when the initial radius $\Delta_0$ is prescribed. Each outer iteration fixes $x$, $J$, and $F$, then tries a sequence of damping values. Every trial changes the diagonal $\mu D$, refactorizes the same augmented pattern, and computes a candidate step. This is the nested work counted in our experiments.

If a candidate exceeds $\Delta$, it is not evaluated; $\mu$ is doubled and the step is recomputed. Otherwise, the gain ratio $\rho$ in~\eqref{eq:gain_ratio} is evaluated. A step with $\rho>10^{-4}$ is accepted. An evaluated rejection halves $\Delta$ and doubles $\mu$. After an accepted step, the radius is updated as
\begin{equation}
\Delta^{+}=
\begin{cases}
\Delta/2, & \rho<0.25,\\
2\Delta, & \rho>0.75\ \text{and}\ \|\delta\|_D>\Delta/2,\\
\Delta, & \text{otherwise},
\end{cases}
\label{eq:tr_update}
\end{equation}
with the prescribed minimum and maximum radii enforced. The damping is divided by three when $\rho>0.75$ and otherwise left unchanged. Thus $\mu$ shapes the trial step, while $\Delta$ imposes an independent bound on its length.

\subsection{Three-Step Composite LM}
We use \textbf{TS} for our residual-scaled three-step LM implementation~\cite{derakhshandeh2018}. Each outer iteration evaluates three corrections from the same $x$, $J$, and $F$ through one factorization and two additional triangular resolves. The fresh-$J$ control reassembles and refactorizes at every correction. Published TSLM and FTSLM results retain their original names in the literature comparison.

\subsection{Adaptive TR: One-Probe Initialization}
Adaptive TR changes only the initialization of TR. Its accepted-step test, damping updates, and radius updates are identical to those above. The extension addresses two opposite failures of a prescribed $\Delta_0$: an unnecessarily small radius slows a healthy start, while an excessively large radius admits a damaging first step from a poor flat start.

At $x_0$, compute one probe direction $d$ with $\mu_p=10^{-6}$ and evaluate the merit retention
\begin{equation}
    r=\frac{f(x_0+d)}{f(x_0)}.
    \label{eq:retention}
\end{equation}
If $r\leq\kappa$, the probe indicates a trustworthy local model: the radius is retired and the solve continues as damped Newton. If $r>\kappa$, adaptive TR initializes
\begin{equation}
    \Delta_0=\|d\|_D\sqrt{\frac{\kappa}{r}},
    \qquad
    \mu_0=\min\{\max(\mu_{\mathrm{init}},r),\mu_{\max}\},
    \label{eq:adaptive_initial}
\end{equation}
clipping $\Delta_0$ to its configured bounds, and then enters the TR loop. The initialization probe costs one factorization and one residual evaluation per solve and is excluded from accepted-step counts. This retention-based joint initialization is our extension of classical trust-region LM~\cite{nocedal2006}.

In the tested systems, healthy starts gave $r=0.003$--$0.14$, while diverging flat starts on stressed systems gave $r=200$--$10^4$. We use $\kappa=5$ for all reported adaptive-TR runs.

\subsection{Assembly-Path Terminology}
We use \textbf{NE} for the damped normal equations and \textbf{AUG} for the augmented system. \textbf{FS} constructs the full Jacobian before reduction, \textbf{COO} uses coordinate-list assembly, and \textbf{SDF} means symbolic direct fill. \textbf{TRIU} stores only the upper triangle. Table~\ref{tab:assembly_paths} defines every path used in the assembly comparison.

\begin{table}[t]
\centering
\caption{Assembly paths and their algorithm types.}
\label{tab:assembly_paths}
\setlength{\tabcolsep}{4pt}
\scriptsize
\begin{tabularx}{\columnwidth}{@{}lX@{}}
\toprule
Path & Construction and structural reuse\\
\midrule
NE-cached & Reduced $J$; fixed-pattern $J^{\top}J$; numeric sparse product at every iteration\\
NE-rebuild & Reduced $J$; generic sparse product rebuilds the $J^{\top}J$ structure and values\\
AUG-FS & Full $J$ followed by slicing; full augmented COO assembly\\
AUG-FS-upper & Full $J$ followed by slicing; upper-triangle augmented COO assembly\\
AUG-COO & Reduced $J$; full augmented COO assembly\\
AUG-COO-upper & Reduced $J$; upper-triangle augmented COO assembly\\
AUG-upper & Direct upper-triangle fill using stored row and column offsets\\
AUG-operator & Proposed direct upper-triangle fill; addresses derived from $\Ybus$ at loop entry\\
\bottomrule
\end{tabularx}
\end{table}

Every COO path inserts triplets and converts them to CSC during each linear solve. The two NE paths form $J^{\top}J+\mu D$ explicitly. AUG-upper and AUG-operator fill the augmented matrix directly. The main comparison uses AUG-operator, the fastest COO control, and NE-cached; the remaining paths provide ablations.

\subsection{Experimental Controls}
We measure optimized release builds on a Windows host with an Intel Core i7-10700K processor at 3.80~GHz. Assembly variants execute identical iterates and share the same Jacobian kernels. Backend comparisons supply QDLDL and KLU with the same full augmented matrix at every iterate and verify equal iteration and factorization counts. Assembly and backend timings are medians of seven rounds after one warmup. Algorithm-level wall-clock tables report one complete optimized-release run per method, including all accepted and rejected trials. Initialization and symbolic analysis are excluded from the steady-state backend comparison. For QDLDL, the numeric timer covers value updates and refactorization on every linear solve. For KLU, value-only refactorization is timed separately from the first factorization, which is excluded; each round's refactorization total is therefore divided by its refactorization count. Triangular-solve totals are divided by the number of right-hand sides for both backends. We normalize within each round before taking cross-round medians.

Convergence tests use flat and dc starts with identical network data and tolerances. Loading scans scale the specified net injections along one fixed direction. Residual and merit identify solved points and nonzero-residual stationary points; Jacobian conditioning is evaluated at a specified solution; continuation brackets the fold along the same loading direction. These measurements distinguish initialization failure, solution existence, and Jacobian conditioning.

\section{Numerical Results}
\label{sec:results}

\subsection{Assembly and End-to-End Speedups}
\begin{figure*}[t]
\centering
\includegraphics[width=\textwidth]{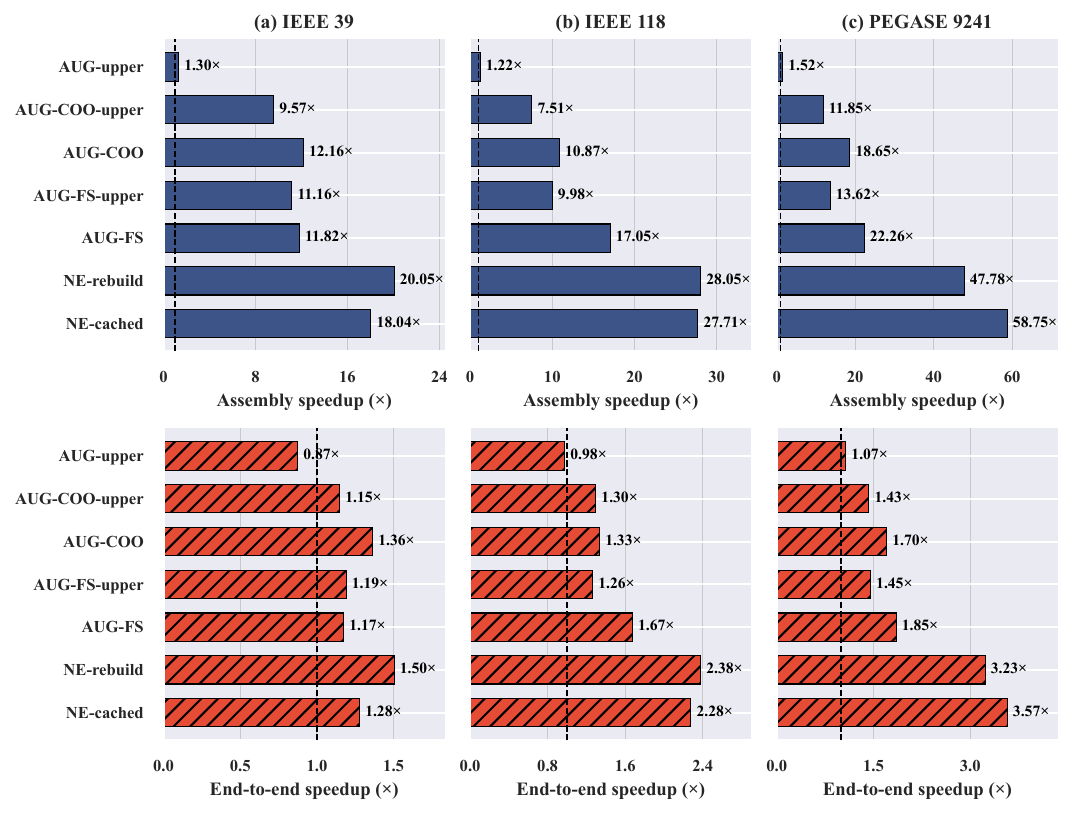}
\caption{Assembly and end-to-end speedups of AUG-operator over every alternative path.}
\label{fig:assembly_speedups}
\end{figure*}

Table~\ref{tab:assembly} and Fig.~\ref{fig:assembly_speedups} report the speedup of AUG-operator over every alternative path. Each ratio divides the alternative path's time by the AUG-operator time, so values above one favor AUG-operator. All ratios use complete LM solves with every damping trial and factorization included. All paths report $4/4$, $5/5$, and $11/11$ accepted steps/factorizations on the three systems.

\begin{table*}[t]
\centering
\caption{Speedup of AUG-operator over each path; higher values are better.}
\label{tab:assembly}
\setlength{\tabcolsep}{7pt}
\begin{tabular*}{\textwidth}{@{\extracolsep{\fill}}lrrrrrr@{}}
\toprule
& \multicolumn{2}{c}{IEEE 39} & \multicolumn{2}{c}{IEEE 118} & \multicolumn{2}{c}{PEGASE 9241}\\
\cmidrule(lr){2-3}\cmidrule(lr){4-5}\cmidrule(lr){6-7}
Path & Assembly & End-to-end & Assembly & End-to-end & Assembly & End-to-end\\
\midrule
AUG-operator & 1.00 & 1.00 & 1.00 & 1.00 & 1.00 & 1.00\\
AUG-upper & 1.30 & 0.87 & 1.22 & 0.98 & 1.52 & 1.07\\
AUG-COO-upper & 9.57 & 1.15 & 7.51 & 1.30 & 11.85 & 1.43\\
AUG-COO & 12.16 & 1.36 & 10.87 & 1.33 & 18.65 & 1.70\\
AUG-FS-upper & 11.16 & 1.19 & 9.98 & 1.26 & 13.62 & 1.45\\
AUG-FS & 11.82 & 1.17 & 17.05 & 1.67 & 22.26 & 1.85\\
NE-rebuild & 20.05 & 1.50 & 28.05 & 2.38 & 47.78 & 3.23\\
NE-cached & 18.04 & 1.28 & 27.71 & 2.28 & 58.75 & 3.57\\
\bottomrule
\end{tabular*}
\end{table*}

The assembly speedup reaches $7.5$--$11.9\times$ over the fastest COO path and $18.0$--$58.8\times$ over NE-cached. The corresponding end-to-end speedups are $1.15\times$, $1.30\times$, and $1.43\times$ over the fastest COO path, and $1.28\times$, $2.28\times$, and $3.57\times$ over NE-cached. AUG-upper has the lowest end-to-end time on IEEE~39 and IEEE~118; AUG-operator has the lowest end-to-end time on PEGASE~9241.

Correctness checks compare matrices and complete trajectories. The row-oriented baseline reproduces the production column Jacobian bitwise on IEEE~39. All construction paths produce identical accepted-step and factorization counts, and backend solutions agree to floating-point precision. On PEGASE~9241, the QDLDL and KLU augmented paths agree to $\max|\Delta V|=6.2\times10^{-15}$ in the recorded cross-check.

\subsection{Factorization Backend on Identical Matrices}
\begin{figure*}[t]
\centering
\includegraphics[width=\textwidth]{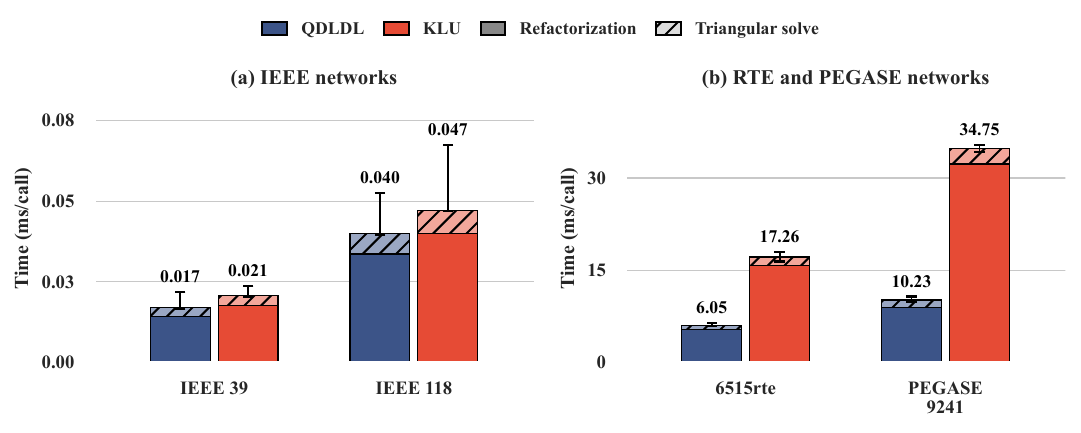}
\caption{QDLDL and KLU cost on the same augmented matrices.}
\label{fig:backend_phases}
\end{figure*}

We compare steady-state numerical work separately from initialization. Fig.~\ref{fig:backend_phases} reports numerical refactorization and triangular solves, with their sum per linear solve printed above each bar. Both backends receive the same augmented matrices on identical trajectories. QDLDL is faster on every system, reaching a KLU/QDLDL ratio of $3.40\times$ on PEGASE~9241. The 6515rte comparison uses dc-started iterates.

The normal-equation KLU control exposes a separate correctness issue. On flat-started case6515rte, value-only refactorization failed in every measured round and stalled at $\|F\|_\infty\approx1.509\times10^3$, with zero factorization-failure fallbacks reported. Fresh KLU factorization at every trial converged along the reference trajectory with 24 accepted iterations and 44 full factorizations, requiring 2.242~s of linear-solve time. The successful KLU path therefore required 44 full factorizations, in contrast to the attempted pattern of one full factorization and 43 value-only refactorizations. The augmented QDLDL path reused its fixed symbolic ordering and required 0.312~s on the same accepted-step sequence. This reliability requirement makes QDLDL central to the design: KLU incurs both its higher large-system cost and the cost of fresh full factorization when pivot reuse fails.

\subsection{RTE Systems: Initialization Versus Existence}
Table~\ref{tab:rte} reports the three stressed RTE snapshots. Flat-start NR fails at base loading on all three, whereas dc-started NR converges in 6--8 iterations. Throughout the loading scan, dc+NR converges at every tested point before the continuation fold. Its first failure occurs after the loading boundary is crossed.

\begin{table*}[t]
\centering
\caption{Dual-start convergence and loading limits of the RTE systems.}
\label{tab:rte}
\setlength{\tabcolsep}{6pt}
\begin{tabular*}{\textwidth}{@{\extracolsep{\fill}}lcccc@{}}
\toprule
Case & Flat NR at $\alpha=1$ & dc+NR iterations & Highest solved dc+NR bracket & Continuation fold\\
\midrule
case6470rte & Fails & 6--7 & $(1.20,\,1.25]$ & 1.2426\\
case6495rte & Fails & 7 & $(1.15,\,1.20]$ & 1.1669\\
case6515rte & Fails & 7--8 & $(1.10,\,1.15]$ & 1.1058\\
\bottomrule
\end{tabular*}
\end{table*}

The dc angle correction at base loading is $0.05$--$0.08$ radians, yet it moves the first iterate into the NR convergence basin. Flat-start failure therefore measures starting-point sensitivity in these systems; the condition numbers reported below characterize the equations at converged solutions.

\subsection{Case2383wp: Conditioning and Reported Cost}
Prior work reported NR divergence and a condition number of $2.3\times10^{12}$ for case2383wp~\cite{derakhshandeh2018}. In our implementation, NR converges in four iterations from both flat and dc starts, so the base case has a readily reachable power-flow solution. The historical parameter settings in the MATPOWER case definition explain the reported divergence. Both tested phase-shifter sign conventions converge. The estimated condition number is $6.0\times10^5$ at the base-case solution and $2.8\times10^6$ near the continuation fold at $\alpha\approx1.88$. The genuine near-singularity appears near the fold.

The original TS study reports outer iterations, while the nested factorizations and additional resolves remain unreported~\cite{derakhshandeh2018}. Table~\ref{tab:case2383} shows why this distinction matters. Our TS requires 84 outer iterations. Every iteration performs exactly one factorization and two additional triangular resolves, giving 84 factorizations and 168 resolves in 237~ms. Reassembling the Jacobian at every correction also gives 84 iterations. Adaptive TR requires 9 iterations and 10 factorizations in 30.3~ms, so it remains faster in our implementation.

Published TSLM and FTSLM report 11 and 5 outer iterations with CPU times of 441.72~s and 200.780~s. Our TS executes more outer iterations yet runs $1.86\times10^3$ and $847$ times faster than these reported times. Direct Jacobian assembly contributes through the gains in Table~\ref{tab:assembly}. The timing ratios also include implementation and hardware effects, although processor generation alone is unlikely to explain differences of this size. Complete factorization, resolve, assembly, and solver measurements therefore provide the basis for performance comparisons.

\begin{table}[t]
\centering
\caption{Case2383wp cost. TS entries separate factorizations and additional resolves.}
\label{tab:case2383}
\setlength{\tabcolsep}{3pt}
\begin{tabular*}{\columnwidth}{@{\extracolsep{\fill}}lrrr@{}}
\toprule
Method & Outer iter. & Fact. + resolves & Time\\
\midrule
Published TSLM & 11 & --- & 441.72~s\\
Published FTSLM & 5 & --- & 200.780~s\\
Our TS, flat & 84 & $84+168$ & 237~ms\\
Our adaptive TR, flat & 9 & 10 & 30.3~ms\\
Our NR, flat & 4 & 4 & 7.58~ms\\
\bottomrule
\end{tabular*}
\end{table}

\subsection{Tripathy 11-Bus System: A Nonzero Merit Floor}
The 11-bus system of Tripathy \emph{et al.}~\cite{tripathy1982} exhibits a different condition. Continuation locates the fold at $\alpha\approx0.643$, below the reported nominal loading. LM converges from both starts to the same least-squares stationary point with
\begin{equation}
    f(x^\star)=\tfrac12\|F(x^\star)\|_2^2
    =8.917\times10^{-4}.
\end{equation}
Together with the continuation result, the nonzero residual places the nominal specification beyond the feasible branch. These findings support Iwamoto and Tamura's assessment of this case~\cite{iwamoto1981}. LM returns the least-squares point and its mismatch distribution for the unattainable operating request.

\begin{table*}[!t]
\centering
\caption{RTE iterations, linear work, and wall time (ms).}
\label{tab:variants}
\setlength{\tabcolsep}{4pt}
\begin{tabular*}{\textwidth}{@{\extracolsep{\fill}}lcccc@{}}
\toprule
Case & dc+NR & TS, flat & TS fresh-$J$, flat & Adaptive TR, flat\\
\midrule
case6470rte & 6/6; 30.2 & 152/(152+304); 1545 & 150/451; 3419 & \textbf{24/25; 207}\\
case6495rte & 7/7; 31.7 & 175/(175+350); 1770 & 175/525; 3945 & \textbf{18/19; 162}\\
case6515rte & 8/8; 39.8 & 245/(245+490); 2731 & 242/726; 5534 & \textbf{22/23; 235}\\
\bottomrule
\end{tabular*}
\end{table*}

\subsection{LM Variants Under Factorization Accounting}
Table~\ref{tab:variants} compares residual-scaled composite LM, fresh-Jacobian composite LM, and adaptive TR. Entries give outer iterations/factorizations, with additional resolves in parentheses for TS. TR and adaptive TR share the trial loop in Section~\ref{sec:protocol}: TR starts from a prescribed radius, whereas adaptive TR initializes the radius and damping with one probe. A composite outer step evaluates three corrections through one shared factorization, followed by triangular resolves. The fresh-Jacobian control reassembles and refactorizes at every correction.

The adaptive trust-region variant converges from the RTE flat starts in 19--25 total factorizations, including the initial probe. Relative to TS, its wall-time speedups are $7.5\times$, $10.9\times$, and $11.6\times$; relative to fresh-Jacobian TS, they are $16.5\times$, $24.3\times$, and $23.6\times$. On healthy systems it retires the radius and behaves as damped Newton with one additional probe factorization: the iteration/factorization counts are $4/5$, $5/6$, and $8/9$ on IEEE 39, IEEE 118, and IEEE 300. The best hand-tuned TR on case6470rte still requires 44--48 factorizations.

Residual-scaled composite LM converges without parameter tuning but requires 152--245 factorizations. Fresh-Jacobian corrections change the accepted-iteration count by less than $2\%$, so Jacobian reuse leaves the long trajectories essentially unchanged. Scalar residual-dependent damping couples step length and direction, whereas the trust region controls them separately.

The published outer-iteration metric misses this distinction. In one case6495rte run, a single-step residual-scaled variant reports 15 outer iterations while executing 1{,}214 factorizations through damping retries. An outer-loop ranking can therefore assign an advantage to an LM variant with orders of magnitude more factorization work.

\section{Discussion}
\label{sec:discussion}

\subsection{A Reproducible Meaning of Ill-Conditioned Power Flow}
The RTE and case2383wp experiments give flat-start failure a precise but limited meaning: it records the outcome from one initial point. Controlled dc initialization and corrected case parameters can change that outcome. Jacobian conditioning must therefore be evaluated at a stated solution and loading point.

A reproducible ill-conditioned benchmark should provide the exact network version and parameters, loading direction, initialization, and convergence tolerance. A dc+NR control and a continuation-based loading boundary separate starting-point sensitivity from loss of solution existence. Factorization and resolve counts complete the performance record. These elements belong in the benchmark definition.

\subsection{The Appropriate Role of LM Power Flow}
Trust-region LM is useful for difficult flat starts. Adaptive initialization chooses the starting radius and damping, while subsequent radius updates continue to shape the accepted path and computational cost. NR remains the simpler method whenever dc+NR converges.

Beyond the feasible operating region, LM returns a least-squares stationary point of~\eqref{eq:merit}. Its merit, gradient, and bus-wise mismatches quantify the unattainable operating request. Continuation locates the loading boundary, and the LM residual describes the outcome beyond it.

\section{Conclusion}
\label{sec:conclusion}
This paper presented an augmented LM system with exact normal-equation equivalence, proven quasi-definiteness, and a symbolic direct-fill implementation. The pre-permuted $\Ybus$ structure supplies every column base, diagonal rank, and segment offset during assembly. AUG-operator achieves $7.5$--$11.9\times$ assembly speedups over the fastest COO control and $18.0$--$58.8\times$ over cached normal-equation sparse-product formation. The respective end-to-end speedups are $1.15$--$1.43\times$ and $1.28$--$3.57\times$. QDLDL reduces numerical refactorization plus triangular-solve time by up to $3.40\times$ on identical augmented systems. Its fixed-order factorization also avoids the fresh full KLU factorizations required when pivot reuse fails on LMPF trajectories.

Complete linear-algebra accounting changes the performance comparison. On case2383wp, 84 TS iterations with 84 factorizations and 168 additional resolves take 237~ms. The contrast with published runtimes shows that outer-iteration counts give an incomplete measure of implementation cost. Our one-probe adaptive trust-region extension further reduces the flat-started RTE solves to 19--25 factorizations and runs $7.5$--$11.6\times$ faster than the reproduced TS method.

The initialization experiments provide equally important context. The dc+NR method converges at every tested RTE loading point before the continuation fold, and NR solves case2383wp in four iterations. Reproducible benchmarks should therefore report case parameters, initialization, loading, tolerances, factorizations, and resolves. Trust-region LM remains useful for difficult flat starts, with its radius influencing convergence and cost. Its clearest practical role is least-squares analysis beyond the feasible operating region.

\bibliographystyle{IEEEtran}
\bibliography{bib/lmpf}

\end{document}